\documentstyle[12pt,epsfig]{article}
\newlength{\dinwidth}
\newlength{\dinmargin}
\newcommand{\ba}{\begin{array}}
\newcommand{\ea}{\end{array}}
\newcommand{\be}{\begin{equation}}
\newcommand{\ee}{\end{equation}}
\newcommand{\bea}{\begin{eqnarray}}
\newcommand{\eea}{\end{eqnarray}}
\newcommand{\gsim}{\mathrel{\mathop{\kern 0pt \rlap
  {\raise.2ex\hbox{$>$}}} \lower.9ex\hbox{\kern-.190em $\sim$}}}

\def\ben{\begin{equation}}
\def\een{\end{equation}}
\def\bea{\begin{eqnarray}}
\def\eea{\end{eqnarray}}
\def\nn{\nonumber}

\begin{document}
\thispagestyle{empty}
\addtocounter{page}{-1}
\vskip-0.35cm
\begin{flushright}
UK-04-18 \\
CERN-PH-TH/2004-147 \\
\end{flushright}
\vspace*{0.2cm}
\centerline{\Large \bf Open-closed duality and Double Scaling}
\vspace*{1.0cm} 
\centerline{{\bf Sumit R. Das${}^{a}$} and  {{\bf Cesar Gomez${}^b$}}}
\vspace*{0.7cm}
\centerline{\it Department of Physics and Astronomy,}
\vspace*{0.2cm}
\centerline{\it University of Kentucky, Lexington, KY 40506 \rm USA ${}^a$}
\vspace*{0.35cm}
\centerline{\it Instituto de Fisica Teorica, C-XVI Universidad Autonoma,}
\vspace*{0.2cm}
\centerline{\it E-28049 Madrid \rm SPAIN ${}^b$}
\vspace*{0.35cm}
\centerline{\it Theory Division, CERN, Geneva, CH-1211 \rm SWITZERLAND ${}^b$}
\vspace*{1cm}
\centerline{\tt das@gravity.pa.uky.edu
\hskip0.7cm Cesar.Gomez@uam.es}

\vspace*{0.8cm}
\centerline{\bf abstract}
\vspace*{0.3cm}
Nonperturbative terms in the free energy of Chern-Simons gauge
theory play a key role in its duality to the closed topological string. We
show that these terms are reproduced 
by performing a double scaling limit near the
point where the perturbation expansion diverges. This leads to
a derivation
of closed string theory from this large-N gauge theory 
along the lines of noncritical string theories.
We comment on the
possible relevance of this observation to the derivation of 
superpotentials of asymptotically free gauge theories and its relation
to infrared renormalons.
\vspace*{0.5cm}

\baselineskip=18pt
\newpage

\section{Introduction}

Ever since 't Hooft's seminal work \cite{thooft}
it has been known that field
theories in the large-N limit become {\em closed} string theories,
with ${1\over N}$ playing the role of string coupling constant. Recent
developments have revealed that the emergence of such closed string
theories is a manifestation of open-closed duality and several
well understood examples of such dualities, e.g. the duality of noncritical
string theory in two spacetime dimension with matrix quantum mechanics
\cite{ceqonea} and
the duality of certain topological open string theories with topological
closed string theories \cite{topological}-\cite{Diaconescu:2002sf}. 
The $AdS/CFT$ correspondence \cite{adscft} is also of this
class, though we do not completely understand the formulation of the bulk
string theory in this case.

Such dualities have provided us important information about long
standing problems in gauge theories. Specifically, in asymptotically
free gauge theories like those with $N=1$ supersymmetry,
nonperturbative contributions to the effective superpotential
\cite{vy} have a
direct connection with the topological version of open-closed duality
\cite{topovy}.
In this paper we initiate a slightly different understanding of these
nonperturbative terms based on a double scaling limit
\cite{dscaling} of the 
underlying large-N gauge theory.

In a typical large-N matrix model characterized by a coupling constant
$g$, usual perturbation theory has a {\em finite} radius of
convergence at $g=g_c$ \cite{bipz}.  
At this point the average number of vertices
in a typical Feynman diagram diverges so that the diagram becomes a
continuous two dimensional surface. If the singularity has the same
location $g_{c}$ at every genus order in the $\frac{1}{N}$ expansion
we can define, by performing a double scaling limit, a continuum non
critical closed string theory.  Generically the contribution to the
free energy at genus $g$ diverges as
\ben
F_g \sim (g-g_c)^{\chi(1+\frac{1}{2m})}~~~~~~~~~~g \geq 2
\label{one}
\een
where $\chi= 2-2g$ is the Euler characteristic and the integer 
$m$ depends on the particular matrix model. 
This behavior allows a definition of
nonperturbative string theory by passing to the double scaling limit
\bea
& & N  \rightarrow  \infty~~~~~~g \rightarrow g_c \nn \\
& & N(g-g_c)^{-(1+\frac{1}{2m})}  =  \mu = {\rm finite}
\label{two}
\eea
In this string theory genus $g$ amplitudes depend on $\mu$ as $\mu^{\chi}$
and various quantities are functions of the single parameter 
$\mu$ rather than on $N$ and $g$ separately. The emergence of a single
parameter is in fact a hallmark of {\em noncritical string theory}. 
This simply means that in this string theory, the constant part of the 
value of the dilaton
is not independent of the values of other backgrounds, as opposed to 
critical string theory where the dilaton is a modulus and can have an 
arbitrary value. In the double scaling limit the coupling constant $g$ 
is renormalized to
\ben
g_{R}
=\frac{g-g_{c}}{a^{2}}
\een
and
the {\em bare} 
string coupling constant $\frac{1}{N}$ is renormalized to 
\ben
g_{s}^{R}=\frac{a^{-(2+\frac{1}{m})}}{N}
\een
$a$ has dimensions of length and plays the role of a cutoff
on the random surface. The parameter
$\mu$ defined above is then the dimensionless ratio
\ben
\frac{g_{R}^{-(2+\frac{1}{m})}}{g_{s}^{R}}
\een

The emergence of closed string theories in the $AdS/CFT$ correspondence
or in topological string theory appears to be quite different at first
sight. For example in 
the $AdS_5/CFT_4$ correspondence, the open string theory in fact becomes
a $SU(N)$ gauge theory characterized by a coupling constant $g_{YM}$. The 
dual closed string theory also has two independent parameters : the string
coupling which is 
\ben
g_s^{crit} = g_{YM}^2
\label{four}
\een
 and the $AdS$ scale in string units is,
\ben
R/l_s \sim (g_{YM}^2 N)^{1/4},
\een
as appropriate for a critical string theory.
Similarly in the simplest topological context, the theory of matrices is
a Chern-Simons gauge theory while the dual theory is a closed topological
string living on a target space characterized by a parameter $t$ (the 
complexified area of the $S^2$ resolution of the conifold in string units,
whose imaginary part is the B-field on the $S^2$) 
and a string
coupling $g_s$ where once again $g_s = g_{YM}^2$ and $t = ig_{YM}^2 N$.
In this latter case, the duality is exactly known and {\em
nonperturbative} terms of the Chern Simons free energy play an
essential role \cite{topological,oogurivafa}.
 
One might wonder one can construct a closed string theory
similar to noncritical strings starting from Chern Simons theory, 
using double scaling limit at 
the radius of convergence of the {\em perturbative} expansion
In this paper we show
that this is indeed true and that the double scaled theory
exactly corresponds to the $c=1$ model ($m=\infty$) at self dual
radius. The latter is known to be equivalent to closed topological
(B model) strings on a $S^3$ deformation of the conifold
\cite{ghoshal}.
Furthermore we show that the free energy of the double scaled
theory precisely reproduces the {\em nonperturbative} terms 
of Chern Simons theory by using level
rank duality. 

The ability to reproduce nonperturbative terms from the perturbative
expansion using double scaling tempts us to conjecture that a similar
procedure could be valid for other theories as well. In particular we
point out that the structure of perturbation theory for Chern-Simons
is exactly that of the Borel transform of perturbation expansion of
asymptotically free theories and the point beyond which the perturbation
expansion diverges
corresponds to the first infrared renormalon singularity. It
has been recently argued in \cite{gomez} that these renormalons are
key ingredients in the generation of mass gap in asymptotically
free theories.
This
connection may well lead to ways of computing nonperturbative terms
in asymptotically free theories.

\section{Chern-Simons Free energy and toplogical strings}

Open topological string on $T^*S^3$ is exactly three dimensional Chern
Simons theory \cite{witten}. 
In the large-$N$ expansion, the free energy has an expansion
\ben
F = \sum_{g=0}^\infty N^{2-2g}F_g (\lambda)
\label{five}
\een
where
\ben
\lambda = g_{YM}^2 N = {2\pi N \over k+N}
\label{six}
\een
is the 't Hooft coupling and $k$ is the level of the Kac-Moody algebra.
The genus $g$ term $F_g$
has a perturbative contribution $F^{p}_g$ and a non-perturbative
contribution $F^{np}_g$ where \cite{topological}
\bea
F^p_0 (\lambda) & = & 2 \sum_{p=2}^\infty
{\zeta(2p-2) \over 2p(2p-1)(2p-2)}({\lambda \over 2\pi})^{2p-2} \nn \\
F^p_1 (\lambda) & = &  \sum_{p=1}^\infty B_{2} {\zeta (2p)
\over 2p}  ({\lambda \over 2\pi})^{2p} \nn \\
F^p_g (\lambda) & = & 2 \chi_{g}\sum_{p=1}^\infty \zeta(2g-2+2p)~
\left(
\begin{array} c
{2g-3+2p} \\ 
{2p} 
\end{array}
\right)
\left( {\lambda \over 2\pi} \right)^{2g+2p-2}~~~~~~~~~g \geq 2
\label{seven}
\eea
and \cite{oogurivafa}
\bea
F^{np}_0 (\lambda) & = & {1\over 2}N^2(\log~(2\pi i\lambda)
 -{3\over 2}) \nn \\
F^{np}_1 (\lambda) & = & -{1\over 12}\log~N \nn \\
F^{np}_g (\lambda) & = & N^{2-2g}\chi_{g}~~~~~~~~~~g \geq 2
\label{sevena}
\eea
where
\ben
\chi_{g} = \frac{B_{2g}}{2g(2g-2)}
\een
and $B_n$ denote the Bernoulli numbers.
These nonperturbative terms arise from the volume of the gauge group
\cite{oogurivafa}
which is nontrivial since the prefactor of the kinetic term is
corrected by quantum effects. They can be obtained in the large $N$ limit using
the asymptotic expansion of the Gamma function.
Apart from $F^{np}_0$ these nonperturbative terms are in fact a function
of $N$ alone.

The expression for the total free energy is in fact exactly that of
A-model topological closed string theory on the $S^2$ resolved conifold
geometry. The string coupling constant $g_s$ is related to the 't Hooft 
coupling by
\ben
g_s = \frac{i\lambda}{N}
\label{eight}
\een
while the complexified Kahler paramter $t$ of the $S^2$ by
\ben
t = i\lambda
\label{nine}
\een
Then the genus-g nonperturbative contribution becomes
\ben
F^{np}_g (\lambda)  =  g_s^{2g-2}{B_{2g}\over 2g(2g-2)~t^{2g-2}}~~~~~~~~~~
g \geq 2
\label{ten}
\een
and similarly for the genus zero and one terms. Thus the closed string
theory is defined at fixed $g_s$ and the nonperturbative terms are
singular in the limit $t \rightarrow 0$, a behavior which is essential
in the identification of the gauge theory with the string theory.

Notice that if we define the Chern Simons string coupling by
$\lambda_{s} = \frac{2\pi}{k+N}$ the last equation in (\ref{sevena})
becomes
\ben
\lambda_{s}^{2g-2} \frac{B_{2g}}{2g(2g-2)\lambda^{2g-2}}
\een
In order to go from this expresion to (\ref{ten}) we need to
replace $\lambda$ by $t=i\lambda$ i.e a Wick rotation in
$\lambda$. The reason for performing this Wick rotation is just in
order to get a positive definite total free energy once we sum over
all genus. In fact the Bernouilli numbers contain a factor
$(-1)^{g-1}$ that makes the total free energy an alternating sum. A
similar phenomena appears in the Penner model
\cite{dv} where a similar Wick
rotation is necessary in order to stablish the connection with the
$c=1$ model at the self dual radius \cite{mukhi}.

\section{A double scaling limit}

We now examine whether a knowledge of the perturbative expansion
can be used to define a critical limit where a continuum string
theory emerges. The perturbative expansions may be summed to 
yield the expressions
\bea
F^p_0 & = & \sum_{n=1}^\infty[{1\over 2}\nu_n^2 \log (\nu_n/N)
-{3\over 4}\nu_n^2 + (n \rightarrow -n)] \nn \\
& & -N^2 \sum_{n=1}^\infty[\log({2\pi \over \lambda})
+ {4\pi^2 n^2 \over \lambda^2}(\log({2\pi \over \lambda})-{3\over 2})
\label{eleven} \\
F^p_1 & = & -{1\over 2}B_2\sum_{n=1}^\infty[\log ({\nu_n \over N})
+ (n \rightarrow -n)] +B_2 \sum_n \log ({\lambda \over 2\pi N}) 
\label{twelveb} \\
F^p_g & = &  \chi_g ({\lambda \over 2\pi})^{2g-2}~
\sum_{n=1}^\infty[ \nu_n^{2-2g} 
+ (n \rightarrow -n)] \nn \\
& & + 2 \chi_g ({\lambda \over 2\pi N})^{2g-2}
~\zeta (2g-2)
\label{elevena}
\eea
where we have defined
\ben
\nu_n = {2\pi N \over \lambda}[{\lambda \over 2\pi} - n]
\label{twelve}
\een

The Chern-Simons theory coupling $\lambda$ has a fundamental domain
between $0$ and $2\pi$. It is clear from the above expressions that
the perturbation expansion has a finite radius of convergence $2\pi$.
In fact the susceptibility diverges at this point with a
characteristic critical exponent. Notice that the divergence of the
perturbative series (\ref{seven}) is for positive value of the Chern
Simons coupling. The reason for this is that only amplitudes with an
even number of holes are non vanishing.

Therefore, following the usual procedure we will define a critical
double scaling limit by taking
\ben
\lambda \rightarrow 2\pi~~~~~~~~~\nu_1 = {\rm finite}
\een
In this limit the nonanalytic terms in the free energy become
\bea
F^{dc}_0 & = & {1\over 2}\nu_1^2 \log (\nu_1 - \frac{3}{2}) \nn \\
F^{dc}_1 & = & -{1\over 12}\log~\nu_1 \nn \\
F^{dc}_g & = & \chi_g \nu_1^{2-2g}
\label{fourteen} 
\eea
where we have used $B_2 = {1\over 6}$.
The coefficients are exactly those of 
the {\em nonperturbative} contributions in 
(\ref{sevena}) with $N$ replaced by $\nu_1$. The string
coupling is renormalized from its bare value ${\lambda \over N}$ to
the renormalized value $g^R_s$ given by 
\ben
g^R_s = {1\over \nu_1}
\label{fifteen}
\een
As is well known this nonanalytic piece is in turn identical to the
free energy of the $c=1$ matrix model at the self-dual radius provided
we Wick rotate $\nu$ to $\mu=i\nu$ as discussed above \footnote{The
connection between the $c=1$ model at the self dual radius and the
singular conifold was done in \cite{ghoshal} where $\mu$ is the
complex modulus.}.

The string theory which is obtained by performing a double scaling limit
near $\lambda = 2\pi$ is in fact related to the string theory defined
using the nonperturbative contribution by level rank duality. This is
given by the interchange
\ben
k \leftrightarrow N
\een
and maps $\lambda =0$ to $\lambda = 2\pi$. Therefore one could read
out the nonperturbative contributions near $\lambda = 0$ from the
double scaled expressions near $\lambda = 2\pi$.  It is important to
notice that the perturbative expansion (\ref{seven}) is perfectly smooth
and analytic at $\lambda=0$, while the non analytic contributions
(\ref{sevena}) are non perturbative and are derived from the
contribution to the Chern Simons free energy of the volume of the
gauge group. Only after including these non perturbative pieces we
recover the closed string picture with the Chern Simons t'Hooft
coupling $\lambda$ related to the size $t$ of the resolved conifold by
$t=i\lambda$. In the double scaling approach, however, we derive the
closed string directly from perturbation theory. This is possible
because the Chern Simons perturbative expansion diverges at
$\lambda=2\pi$. The closed string model that we obtain is the non
critical $c=1$ model at the self dual radius that is known to be
equivalent to topological strings on the deformed conifold
\cite{ghoshal}. 
The double scaled variable
$\nu$ and the deformation parameter $\mu$ of the deformed conifold are
related by 
$\mu = i\nu$. 
In other words, double scaling limit leads to 
closed strings on the deformed conifold i.e the
local mirror version of the topological closed string obtained at
$\lambda=0$. By level rank duality of Chern Simons we can relate both
mirror topological closed string versions one corresponding to the
large level $k$ limit with finite $N$ and the one obtained by double
scaling corresponding to large $N$ and finite $k$.  We would like to
stress that the moral of this exercise is to show that potentially we
can read the dual closed string version of a gauge theory directly
from the divergence structure of the perturbative expansion. In next
section we will present some general comments in this direction.

\section{Structure of perturbation theory, strings and Nonperturbative
Superpotentials}

In the previous section we showed that a double scaling limit of the
{\em perturbative} expansion of Chern Simons theory reproduces the
{\em nonperturbative} contribution to the free energy.

Nonperturbative superpotentials for ${\cal N}=1$ 
$SU(N)$ super Yang Mills can be directly defined using
the nonperturbative piece of the Chern Simons free energy with the same
gauge group. In fact by compactifying on $T^{*}S^{3}$
type A open topological strings we get \cite{BCOV, AGNT} 
perturbative contributions
to the F-term superpotential of type 
\ben
W^{p}(S) = N_{c} \sum F_{0,h} h S^{h-1} = N_{c} \frac{\partial
F^{p}_{0}(S)}{\partial S}
\een
where $F_{0,h}$ are open topological string amplitudes 
at genus zero and with $h$ holes.
Using now the equivalence between open topological strings on
$T^{*}S^{3}$ and Chern Simons on $S^{3}$ one can now define
\ben
W^{np}(S) = N_{c} \frac{\partial F^{np}_{0}(S)}{\partial S}
\een
where $F^{np}_{0}(S)$ is the genus zero non perturbative piece of the
Chern Simons free energy for $\lambda = S$, namely
\ben
F_{0}^{np}(S) = \frac{1}{2} S^{2}(\log~ S - \frac{3}{2})
\een
Notice that the non peturbative contribution to the 
superpotential is defined from $F_{0}^{np}(\lambda)$
given in (\ref{sevena}) 
once we extract the corresponding power $g_{s}^{-2}$ 
of the string coupling constant. In other words
in this approach $S$ is directly related to the Chern Simons coupling
$\lambda$ or in the closed string version to the size $t$ of the
$S^{2}$ resolved conifold.

In the previous section we have derived the non perturbative piece of
Chern Simons free energy by performing a double scaling limit around
the critical coupling defining the radius of convergence of the
perturbative expansion. The result
\ben
F_{0}^{np} = \frac{1}{2} \nu^{2}(\log ~\nu - \frac{3}{2})
\een
Exactly as before we can read the non perturbative superpotential for
$S$ where now $S$ is related with the double scaled variable $\nu$ or
in the closed string version with the complex modulus of the deformed
conifold. This exactly correspond to the mirror type IIB derivation of
the $N=1$ superpotential \cite{topovy}\footnote{As it is
standard in double scaling limit for the $c=1$ model we have
$\nu = \frac{\lambda_{R}}{g_{s}^{R}}$
where $\lambda_{R} = \frac{\lambda - \lambda_{c}}{a^{2}}$ and
$g_{s}^{R}=\frac{1}{a^{2}N}$ for $a$ a worldsheet string scale with
units of length and where $\lambda_{R}$ is defined in the limit $a=0$
and $\lambda= \lambda_{c}$ and $g_{s}^{R}$ is defined in the double
limit $N=\infty$ and $a=0$. In terms of $\lambda_{R}$ and $g_{s}^{R}$
we can write
$F_{0}^{np} = g_{s}^{-2}
\lambda_{R}^{2}(log(\frac{\lambda_{R}}{g_{s}^{R}}) - \frac{3}{2})$
and to extract the nonperturbative piece of $N=1$ super Yang Mills
from $\lambda_{R}^{2}(log(\frac{\lambda_{R}}{g_{s}^{R}}) -
\frac{3}{2})$ using $\lambda_{R}= S$. 
The main difference with the
previous formal construction is that now both $\lambda_{R}$ and
$g_{s}^{R}$ have dimensions.  This on the other hand is quite natural
since $S$ is also dimensionfull.}. 

After our previous analysis it is natural to ask ourselves if we can
use a double scaling limit to define non perturbative physics for
asymptotically free theories. Using t'Hooft's double line notation
the free energy of a generic $SU(N)$ gauge theory has the form
\ben
F= \sum_{g} g_{s}^{2g-2}F_{g}(s)
\een
where $s$ is the t'Hooft coupling 
defined in the large $N$ limit and $g_{s} = \frac {1}{N}$.
Generically $F_{g}(s)$ 
is divergent even for arbitrarily small $s$. 
However it is known that in the large-N limit, 
the Borel transform of $F_{g}(s)$,
denoted by $F^{B}_{g}(\lambda)$ 
has a finite radius of convergence for $\lambda = \lambda_{c}$ where
$\lambda_{c}$ is fixed by the location of the 
first infrared renormalon singularity. Using the Borel transform we can define
\ben
F^{B}(\lambda) = \sum_{g} g_{s}^{2g-2} F_{g}^{B}(\lambda)
\een
of the full free energy. 

We can now try to perform a double scaling limit $\mu = (\lambda -
\lambda_{c})^{\gamma}N$ for some $\gamma$ that will depend on the
concrete theory, and to define a nonperturbative $F^{B}_{np}(\mu)$ in
the manner explained in the previous sections. Of course this strategy
will only work if the critical value fixing the convergency radius
$\lambda_{c}$ is the same at any genus order in the $1/N$
expansion. We can argue that the part of the perturbative expansion
dominated by infrared renormalon singularities satisfy this
criteria.In fact in contrast to instanton singularities
the infrared renormalon is quite universal and does not depend on the
number of diagrams contributing to feynman diagram with some topology
at a given order in the perturbation expansion.

The variable $\lambda$ appearing in the Borel transform can be
naturally interpreted as the conjugated variable to the gauge theory
coupling i.e as a glueball operator and the double scaled non
perturbative Borel transform $F^{B}_{np}(\mu)$ as a candidate for 
the effective infrared low energy physics.

Our observation in the Chern Simons exercise can be rephrased
by saying that the Chern Simons free energy is capturating the
contribution of infrared renormalons to the Borel transform of the
$N=1$ super Yang Mills free energy.  In fact for large number $n$ of
holes the genus zero contribution to the free energy of Chern Simons
goes like 
\ben\label{uno} \frac{N^{2}}{\lambda^{2}}
\sum_{n}\frac{(n-3)!}{n!} (\frac{\lambda}{2\pi})^{n} 
\een If we now consider
(\ref{uno}) as a Borel transform of some perturbative formal series
$F(s)$ in $s$ we get 
\ben\label{tres} F(s) = N^{2}\sum_{n}
\frac{(n-3)!}{n (n-1)} (\frac{s}{2\pi})^{n-1} 
\een If now we formally
think of $s$ as the gauge theory t'Hooft coupling the expansion
\ref{tres} behaves at large $n$ typically as an infrared renormalon
contribution. Recall that for a generic perturbative expansion
$\sum_{n} a_{n} s^{n}$ in t'Hooft coupling $s$ the infrared renormalon
corresponds to $a_{n} \sim b^{n} n!$ for some coefficient $b$ fixed by
the beta function. Notice that the divergence in the perturbative
expansion of the Chern Simons free energy can be interpreted as a
manifestation of the renormalon of $F(s)$ defined in (\ref{tres}).  For
$N=1$ super Yang Mills we find the first renormalon singularity at
$\lambda_{c} = 8\pi^{2}$ while for the Chern Simons analog we get
$\lambda_{c} = 2\pi$.

\section{Acknowledgements} The authors would like to thank 
Rajesh Gopakumar for discussions. S.R.D thanks Instituto de Fisica
Teorica, Universisad Autonoma Madrid, Indian Assiociation for the
Cultivation of Science, Kolkata and Tata Institute of Fundamental
Research, Mumbai for hospitality. This work of S.R.D. is supported by
National Science Foundation grant PHY/0244811 and a Department of
Energy contract DE-FG01-00ER45832. The work of C.G is supported by
Plan Nacional de Altas Energias FBA-2003-02-877.

\end{document}